\documentclass{iau} 
\usepackage{graphicx}
\usepackage{amsmath}
\usepackage{savesym}

\title[IAUS356 -- Accretion and star formation in `radio-quiet' quasars] 
{Accretion and star formation\\ in `radio-quiet' quasars}

\author[Sarah V. White]   
{Sarah V. White (SVW)$^{1,2,3}$, Matt J. Jarvis$^{3, 4}$, Eleni Kalfountzou$^{5,6}$, Martin J. Hardcastle$^{6}$, Aprajita Verma$^{3}$, Jos\'e M. Cao Orjales$^{6}$\\ \and Jason Stevens$^{6}$}

\affiliation{
$^1$South African Radio Astronomy Observatory (SARAO), 2 Fir Street, Observatory, Cape Town, 7925, South Africa.      $^2$Department of Physics and Electronics, Rhodes University, PO Box 94, Grahamstown, 6140, South Africa.      $^{3}$Astrophysics, University of Oxford, Denys Wilkinson Building, Keble Road, Oxford, OX1 3RH, UK \\ email: {\tt sarahwhite.astro@gmail.com} \\[\affilskip]
$^{4}$Department of Physics, University of the Western Cape, Bellville 7535, South Africa\\
$^{5}$European Space Astronomy Centre (ESAC), Villanueva de la Ca\~nada, E-28692 Madrid, Spain\\
$^{6}$Centre for Astrophysics Research, School of Physics, Astronomy and Mathematics, University of Hertfordshire, Hatfield, Herts,
AL10 9AB, UK
}

\pubyear{2019}
\volume{356}  
\jname{Nuclear Activity in Galaxies Across Cosmic Time}
\editors{M. Povic, J. Masegosa, H. Netzer, P. Marziani,\\ P. Shastri, S. B. Tessema \& S. H. Negu, eds.}
\begin{document}

\maketitle

\begin{abstract}
Radio observations allow us to identify a wide range of active galactic nuclei (AGN), which play a significant role in the evolution of galaxies. Amongst AGN at low radio-luminosities is the `radio-quiet' quasar (RQQ) population, but how they contribute to the total radio emission is under debate, with previous studies arguing that it is predominantly through star formation. In this talk, SVW summarised the results of recent papers on RQQs, including the use of far-infrared data to disentangle the radio emission from the AGN and that from star formation. This provides evidence that black-hole accretion, instead, dominates the radio emission in RQQs. In addition, we find that this accretion-related emission is correlated with the optical luminosity of the quasar, whilst a weaker luminosity-dependence is evident for the radio emission connected with star formation. What remains unclear is the process by which this accretion-related emission is produced. Understanding this for RQQs will then allow us to investigate how this type of AGN influences its surroundings. Such studies have important implications for modelling AGN feedback, and for determining the accretion and star-formation histories of the Universe.
\end{abstract}

\firstsection 
\section{Radio emission from `radio-quiet' quasars}

Star formation and black-hole accretion are the key processes that govern how galaxies evolve. Since both processes produce radio emission, deep radio surveys enable complete selection of galaxies out to high redshift, unaffected by dust obscuration (unlike optical surveys) or Compton-thick absorption (unlike X-ray surveys). However, as is the case at other wavelengths, the contributions of star formation and the active galactic nucleus (AGN) towards the total emission need to be disentangled. The techniques we use to do this will prove particularly important for science with the Square Kilometre Array (SKA) and its precursor/pathfinder telescopes (e.g. \cite{McAlpine2015}; \cite{Jarvis2015}). As we probe to lower radio flux-densities ($< 1$\,mJy), a flattening of the source counts implies that we are going from an AGN dominated regime to a star-formation dominated regime (\cite{Condon2012}). However, it is not only star-forming galaxies that are detected below 1\,mJy. We also uncover the population of radio-quiet quasars [RQQs; see predictions by \cite{Wilman2008}], which are a subset of AGN that produce relatively low levels of radio emission when compared to their `radio-loud' counterparts. 

The current assumption is that star formation is responsible for the radio emission from RQQs, with \cite{Kimball2011} and \cite{Condon2013} favouring this interpretation. However, these studies use only radio and optical data for their analyses. For our work, we take a step further by also using far-infrared (FIR) data from the {\it Herschel Space Observatory}, reduced and presented by \cite{Kalfountzou2017}. This allows us to {\it quantify} the level of star formation within our sample of 70 RQQs, by combining the derived far-infrared luminosities with the far-infrared to radio correlation (FIRC). This very tight relation is observed for star-forming galaxies, and arises due to star formation being associated with both dust (producing the FIR emission) and supernova remnants (producing radio emission).

\section{Disentangling black-hole accretion from star formation}

\begin{figure}
\centering
\includegraphics[scale=0.7]{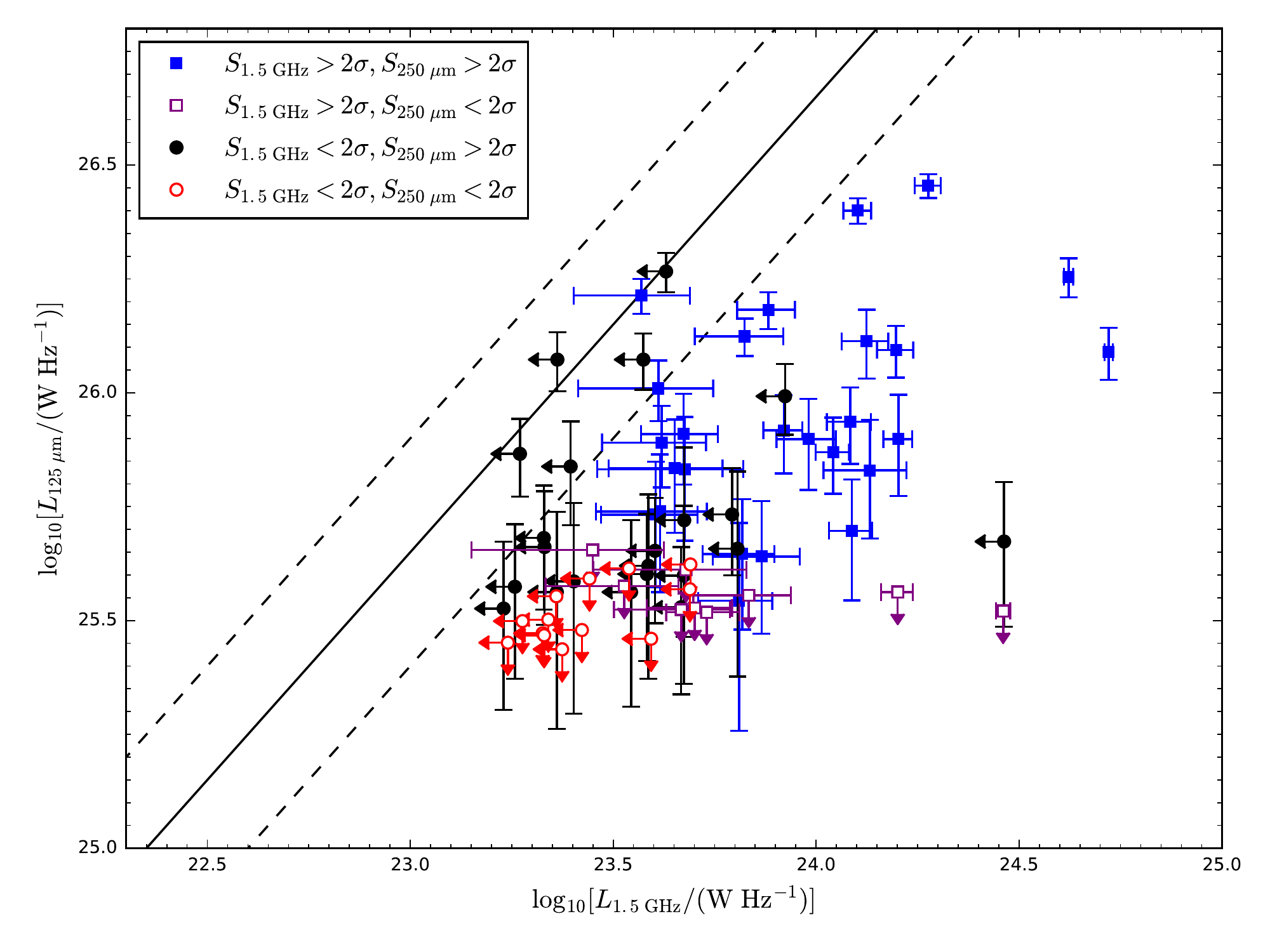} 
\caption{The FIR luminosity at rest-frame 125\,$\mu$m, $L_{125\,\mu{\rm m}}$, versus the radio luminosity, $L_{{\rm 1.5\,GHz}}$. Squares correspond to objects detected in the radio above 2\,$\sigma$, and circles are those below this detection threshold. Note that objects with $\log_{10}[L_{125\,\mu{\rm m}}]\lesssim25.6$ are below 2\,$\sigma$ at 250\,$\mu$m (unfilled symbols). Arrows represent 2-$\sigma$ upper limits in $L_{{\rm 1.5\,GHz}}$ or $L_{125\,\mu{\rm m}}$, for quasars undetected at the 2-$\sigma$ level in the radio (horizontal arrows) or the 250\,$\mu$m photometry (vertical arrows), respectively. The dashed lines are the lower and upper bounds on $q_{125}$ (2.4 and 2.9, respectively), where $q_{125}=\log_{10}[{L_{125\,\mu{\rm m}}}/{L_{{\rm 1.5\,GHz}}}]$. The solid line corresponds to the midpoint value, $q_{125}=2.65$, and represents the FIRC (where star-forming galaxies are expected to lie). {\it Figure reproduced from \cite{White2017}.} }
\label{LradioL125}
\end{figure}

Like the quasars used by \cite{Kimball2011} and \cite{Condon2013}, our sample is selected from the Sloan Digital Sky Survey (SDSS), but we restrict the redshift range to $0.9 < z < 1.1$ in order to minimise any possible evolutionary effects. The radio data we use is in the form of targeted observations using the Karl G. Jansky Very Large Array, enabling us to reach a sensitivity of $\sim30$\,$\mu$Jy in 25 minutes of integration time per source [PI: Jarvis; \cite{White2017}]. When analysed in conjunction with FIR data, we find that for the majority of our RQQs, there is a significant amount of radio emission that cannot be explained by star formation alone. This is illustrated by the RQQs lying to the right of the FIRC in Fig.~\ref{LradioL125}, having `excess' radio emission. The excess emission must be due to another process -- that being black-hole accretion -- and so we use the offsets from the FIRC to calculate the accretion-related radio luminosity. As shown by the numbers in Table~\ref{accretionfraction}, this accretion component is dominating the total radio emission from our RQQs, in support of previous work by \cite{White2015} and going against the assumption that star formation dominates. Our fractions may even be underestimates, given a study by \cite{Wong2016} that shows a sample of hard X-ray selected AGN lying on the FIRC, mimicking the properties of star-forming galaxies. 

\begin{table} 
\centering 
\caption{The accretion-related contribution to the radio luminosity, across our sample of RQQs. An object with $L_{\mathrm{1.5\,GHz,\,acc}}/L_{\mathrm{1.5\,GHz}}>0.5$ is described as being `AGN-dominated'. `Summed radio luminosity' refers to the summation of the total radio luminosity for each object ($\Sigma\,L_{\mathrm{1.5\,GHz}}$), and the `fraction that is accretion-related' is given by $\Sigma\,L_{\mathrm{1.5\,GHz,\,acc}}/\Sigma\,L_{\mathrm{1.5\,GHz}}$. `Upper' and `lower' limits refer to the fraction of the radio emission that is related to accretion, taking into account cases where the object is undetected (i.e. $< 2 \sigma$) in the radio and/or the FIR.} 
\begin{tabular}{@{}lrrrr@{}} 
\hline 
Description of objects used & No. of  & Fraction that are  &  Summed radio & Fraction of summed \\ 
(and the $L_{\mathrm{1.5\,GHz,\,acc}}$ & objects & AGN-dominated & luminosity & luminosity that is  \\
values considered)  &  &  & (W\,Hz$^{-1}$) & accretion-related  \\ 
\hline 
Whole sample  & 70& 0.80 & $3.82\times 10^{25}$ &  0.74 \\ 
Whole sample (lower limits)  &  70 &  $\geq0.47$ &  $\leq5.28\times 10^{25}$  & $\geq0.60$ \\
Whole sample (upper limits) &   70 &  $\leq0.89$ &   $\leq5.28\times 10^{25}$ & $\leq 0.83$ \\
Radio-detected, FIR-detected  &  26 & 0.92 &   $3.07\times 10^{25}$ & 0.80 \\
\hline 
\label{accretionfraction} 
\end{tabular} 
\end{table} 

\begin{figure}
\centering
\includegraphics[scale=0.65]{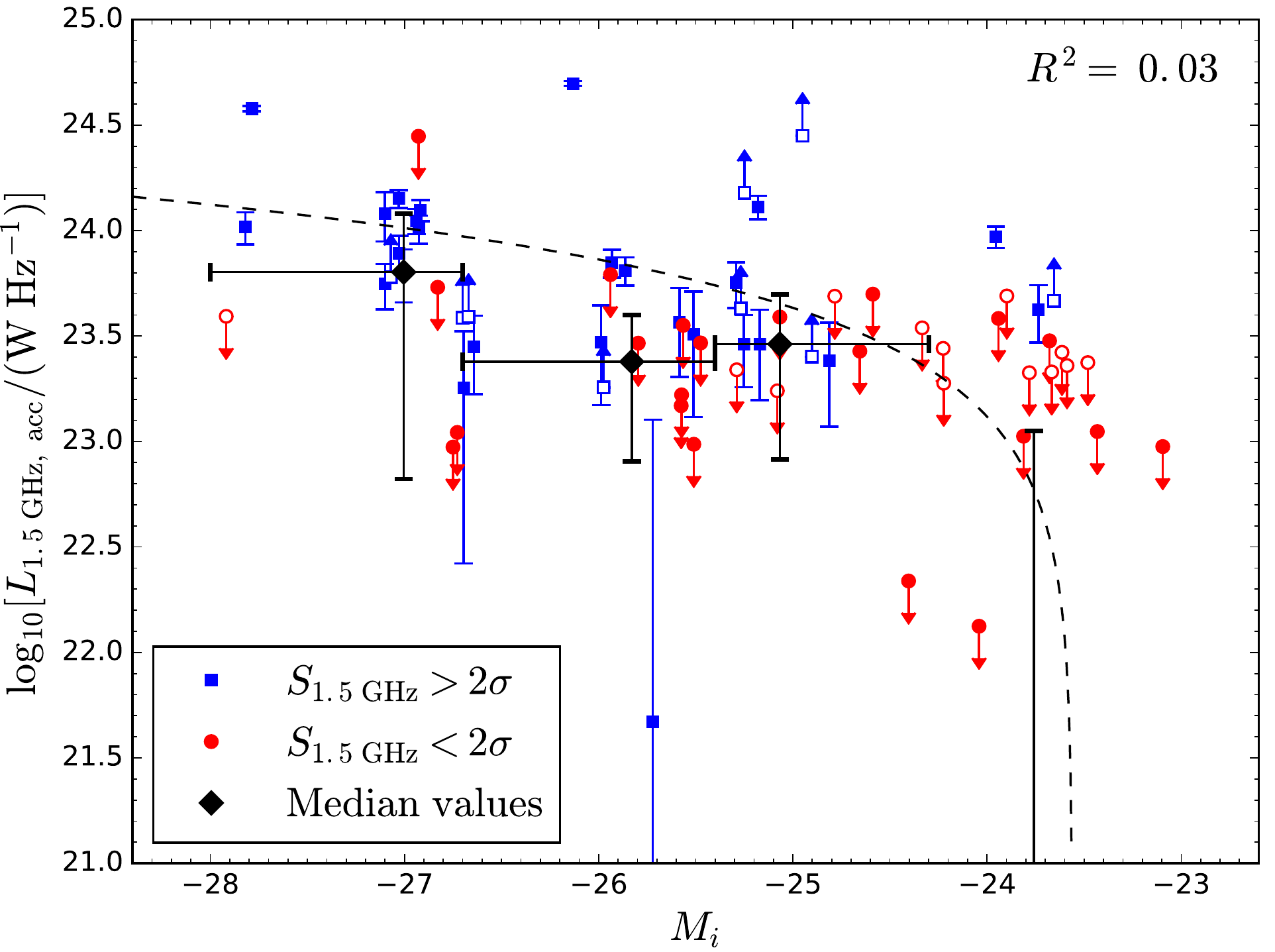}
\caption{The accretion luminosity, $L_{{\rm 1.5\,GHz,\,acc}}$ against the absolute $i$-band magnitude, $M_{i}$. Blue squares correspond to objects detected in the radio above 2\,$\sigma$, and red circles are those below this detection threshold. Unfilled symbols correspond to the FIR data being below 2\,$\sigma$. Arrows indicate whether the value of $L_{{\rm 1.5\,GHz,\,acc}}$ is either an upper or lower limit (at 2\,$\sigma$), dependent on whether the object is undetected in both the radio and the FIR, or undetected in the FIR alone. The line of best-fit is given by $L_{{\rm 1.5\,GHz,\,acc}}=(-2.99\pm 0.84) \times 10^{23} M_{i} - (7.05\pm 2.11)\times 10^{24}$, and the associated coefficient of determination is shown in the top right-hand corner. (Uncertainties in $L_{{\rm 1.5\,GHz,\,SF}}$ and $L_{{\rm 1.5\,GHz,\,acc}}$ are used for this fit.) The dashed line is the result of converting this best-fit line into log--linear space. Overplotted are the median luminosities (black diamonds), derived using all objects, binned in $M_{i}$. The horizontal error-bars indicate the ranges of the $M_{i}$ bins, and uncertainties on the median radio-luminosities are given by the vertical error-bars. Note that the values of the luminosities, even if negative, are used for the linear regression analysis and the calculation of the median luminosities, rather than the limit values. {\it Figure reproduced from \cite{White2017}.} \label{LaccMi} }
\end{figure}

Another aspect of our sample is that it has been downselected in order to enforce a uniform distribution in absolute $i$-band magnitude, spanning a factor of $\sim$100 in optical luminosity. This allows us to explore whether there is any trend in the star-formation-related radio luminosity, or the accretion-related radio luminosity, with the optical luminosity of the quasar. \cite{White2017} find a stronger statistical correlation for the latter combination, which is as expected since the accretion disc of a quasar produces thermal emission that dominates the $i$ band. We therefore have further evidence that this component of the radio emission truly is connected with the AGN. However, a lot of scatter is still seen between the accretion-related radio luminosity and the absolute $i$-band magnitude (Fig.~\ref{LaccMi}), which could be due to variations in magnetic-field strength, environmental density, or the differing timescales over which the radio emission and optical emission are produced.

\section{What accretion-related mechanism is present?}

We have concluded that the AGN is dominating the total radio emission from RQQs, but the exact mechanism involved requires further investigation. It is possible that radio jets (typically associated with `radio-loud' quasars) are being launched (e.g. \cite{Hartley2019}) but they are too small to be resolved in our radio images. Supporting this, earlier in the Symposium we were shown images of small jets in `radio-silent' Seyfert galaxies (\cite{Lahteenmaki2018}), which are believed to be low-redshift analogues of quasars. Another suggestion, regarding the origin of the AGN-related radio emission, is that it is associated with an X-ray corona close to the accretion disc (\cite{Laor2008}), with similar magnetic connection events as seen for coronally-active stars. 

Alternatively, the winds associated with the accretion disc may be impacting upon the surrounding medium and creating shock fronts, which in turn accelerate electrons and generate radio emission. \cite{Hwang2018} provide evidence for this explanation, by looking at the radio luminosity of quasars against the equivalent width of the {\sc [Oiii]} emission line. This quantity acts as a proxy for the quasar's outflow velocity, and they find that an `extreme' sample of high-redshift quasars appears to lie on the same correlation observed for their previous sample of low-redshift quasars. However, they cannot completely rule out the possibility of compact jets being present, as these could give rise to the same kinematics as quasar winds. Indeed, later during the Symposium, \cite{Jarvis2019} presented results for {\it jetted} RQQs lying on the relation by \cite{Hwang2018}.


\section{Implications and future work}

Understanding the origin of the radio emission in radio-quiet AGN is important if we wish to: (i) use faint sources to probe the star-formation history of the Universe, or how accretion activity has evolved; (ii) uncover why some quasars exhibit powerful radio-jets whilst the majority do not; and (iii) see whether we need to consider feedback from radio-quiet AGN as well as from radio-loud AGN. This is particularly important for modelling galaxy evolution. For example, \cite{Mancuso2017} have used our `disentangled' radio luminosities to test their models of star formation and accretion in the radio band, for which they consider in-situ co-evolution of the two processes. They find that our RQQs lie on the average relationship between the accretion radio-luminosity and the star-formation radio-luminosity, averaged across {\it all} galaxies (since radio data offer complete selection).

Most recently, \cite{Malefahlo2019} present radio luminosity functions (RLFs), over multiple redshift ranges, for optically-selected quasars identified in SDSS. For this they use a Bayesian stacking analysis to probe below the 1-mJy flux-limit of the FIRST survey (\cite{Becker1995}). White et al. will test the radio-quiet distribution of these RLFs using optical spectroscopy from the Southern African Large Telescope (PI: White) and deep radio data that has been obtained as part of the MIGHTEE survey (\cite{Jarvis2016}). Following this is the need to disentangle the star-formation and AGN contributions to the total radio emission, for which very long baseline interferometry would offer excellent spatial information to supplement multi-wavelength analysis.


\begin{thebibliography}{}

\bibitem[Becker et al., 1995]{Becker1995}
{Becker, R.H., White, R.L., Helfand, D.J.,} 1995, 
\textit{ApJ}, 450, 559

\bibitem[Condon et al., 2012]{Condon2012}
{Condon, J.J., Cotton, W.D., Fomalont, E.B., Kellermann, K.I., Miller, N., Perley, R.A., Scott, D., Vernstrom, T., \& Wall, J.V.}, 2012, 
\textit{ApJ}, 758, 23

\bibitem[Condon et al. (2013)]{Condon2013}
{Condon, J.J., Kellermann, K.I., Kimball, A.E., Ivezi{\' c}, {\v Z}., Perley, R.A.,} 2013,
\textit{ApJ}, 768, 37

\bibitem[Hartley et al., 2019]{Hartley2019}
{Hartley, P., Jackson, N., Sluse, D., Stacey, H.R., \& Vives-Arias, H.}, 2019,
\textit{MNRAS}, 485, 3009

\bibitem[Hwang et al. (2018)]{Hwang2018}
{Hwang, Hsiang-Chih, Zakamska, N.L., Alexandroff, R.M., Hamann, F., Greene, J.E., Perrotta, S., Richards, \& G.T.,} 2018, 
\textit{MNRAS}, 477, 830

\bibitem[Jarvis et al., 2015]{Jarvis2015}
{Jarvis, M.J., et al.,} 2015,
\textit{PoS}, arXiv:1412.5753

\bibitem[Jarvis et al., 2016]{Jarvis2016}
{Jarvis, M.J., et al.,} 2016,
\textit{PoS}, arXiv:1709.01901

\bibitem[Jarvis et al. (2019)]{Jarvis2019}
{Jarvis, M.E., Harrison, C.M., Thomson, A.P., Circosta, C., Mainieri, V., Alexander, D.M., Edge, A.C., Lansbury, G.B., Molyneux, S.J. \& Mullaney, J.R.,} 2019,
\textit{MNRAS}, 485, 2710

\bibitem[Kalfountzou et al. (2017)]{Kalfountzou2017}
{Kalfountzou, E., Stevens, J.A., Jarvis, M.J., Hardcastle, M.J., Wilner, D., Elvis, M., Page, M.J., Trichas, M., \& Smith, D.J.B.,} 2017,
\textit{MNRAS}, 471, 28

\bibitem[Kimball et al. (2011)]{Kimball2011}
{Kimball, A.E., Kellermann, K.I., Condon, J.J., Ivezi{\'c}, {\v Z}., \& Perley, R.A.,} 2011, 
\textit{ApJ} (Letters), 739, L29

\bibitem[L{\"a}hteenm{\"a}ki et al., 2018]{Lahteenmaki2018}
{L{\"a}hteenm{\"a}ki, A., J{\"a}rvel{\"a}, E., Ramakrishnan, V., Tornikoski, M., Tammi, J., Vera, R.J.C., \& Chamani, W.,} 2018,
\textit{A\&A} (Letters), 614, L1

\bibitem[Laor \& Behar, 2008]{Laor2008}
{Laor, A. \& Behar, E.,} 2008,
\textit{MNRAS}, 390, 847

\bibitem[Malefahlo et al. (submitted)]{Malefahlo2019}
{Malefahlo, E., Santos, M.G., Jarvis, M.J., White, S.V., \& Zwart, J.T.L.,} 2019,
\textit{submitted to MNRAS}, arXiv:1908.05316

\bibitem[Mancuso et al. (2017)]{Mancuso2017}
{Mancuso, C., Lapi, A., Prandoni, I., Obi, I., Gonzalez-Nuevo, J., Perrotta, F., Bressan, A., Celotti, A., \& Danese, L.,} 2017, 
\textit{ApJ}, 842, 95 

\bibitem[McAlpine et al., 2015]{McAlpine2015}
{McAlpine, K., et al.,} 2015, 
\textit{PoS}, arXiv:1412.5771

\bibitem[White et al. (2015)]{White2015}
{White, S.V., Jarvis, M.J., H{\" a}u\ss ler, B., \& Maddox, N.} 2015,
 \textit{MNRAS}, 448, 2665

\bibitem[White et al. (2017)]{White2017}
{White, S.V., Jarvis, M.J., Kalfountzou, E., Hardcastle, M.J., Verma, A., Cao Orjales, J.M., \& Stevens, J.} 2017,
 \textit{MNRAS}, 468, 217

\bibitem[Wilman et al. (2008)]{Wilman2008}
{Wilman, R.J., Miller, L., Jarvis, M.J., Mauch, T., Levrier, F., Abdalla, F.B., Rawlings, S., Kl{\"o}ckner, H.-R., Obreschkow, D., Olteanu, D., \& Young, S.}, 2008,
\textit{MNRAS}, 388, 1335

\bibitem[Wong et al. (2016)]{Wong2016}
{Wong, O.I., Koss, M.J., Schawinski, K., Kapi{\'n}ska, A.D., Lamperti, I., Oh, K., Ricci, C., Berney, S., Trakhtenbrot, B.,} 2016, 
\textit{MNRAS}, 460, 1588

\end{thebibliography}
\end{document}